# Dynamic Model for Query-Document Expansion towards Improving Retrieval Relevance

Onifade F.W. Olufade[1], Arise A. Abiola[2], Ogboo M. Chisom[3]

[1, 2, 3] Department of Computer Science, University of Ibadan, Ibadan, Oyo State, Nigeria
Emails: [1] olufadeo@gmail.com, [2] ariseabiola@yahoo.com, [3] zitaogboo@yahoo.com.

***Abstract***: *Getting relevant information from search engines has been the heart of research works in information retrieval. Query expansion is a retrieval technique which has been studied and proved to yield positive results in relevance. Users are required to express their queries as a short list of words, sentence or questions. With this short format, huge amount of information is lost in the process of translating the information need from the actual query size since the user cannot convey all his thoughts in a few words. This mostly leads to poor query representation which contributes to undesired retrieval effectiveness. This loss of information has made the study of query expansion technique a strong area of study. This research work focuses on two methods of retrieval for both tweet-length queries and sentence-length queries. Two algorithms have been proposed and the implementation is expected to produce a better relevance retrieval model than most state-the-art relevance models.*

## A. INTRODUCTION

With the growing volume of information on the internet, searching for relevant documents from the pool of both relevant and irrelevant information brought about the important need to effectively and efficiently retrieve relevant document within a reasonable time. Information retrieval (IR) involves the activity of getting information resources that are regarded relevant to an information need from a collection of resources. Relevance denotes how well a retrieved document or set of documents meet the information need of the user, it may include concerns such as timeliness, authority or novelty of the result. A search usually begins when a user submits a query to indicate the information need [1]. Most information retrieval systems are built to help reduce the problem of information overload on the users because for most queries, there exist thousands of documents containing some or all of the terms in the query and the IR system needs to rank them in some appropriate way so that the first few results shown to the user will be the ones that are most pertinent to the user's need [2]. Users are required to express their queries as a short list of words, sentence or question. With this short format, huge amount of information is lost in the process of translating the information need from the actual query size since the user cannot convey all his words. This mostly lead to poor query representations which contributes to undesired retrieval effectiveness. This loss of information has made the study of query expansion technique a strong area of study. These techniques are used to supplement the original query to produce a better representation of the user's initial information need. Most previous query expansion techniques represent documents as simple text or bag of words and ignore the relationships and dependencies between document terms.

The importance of a document is determined by assigning a relevance level to each retrieved result through a process known as relevance assessment and thereafter, a ranking function is then employed to rank the documents according to the degree of relevance or importance to the query.

This paper describe and evaluate the design and development of a retrieval model that helps retrieve relevant document depending on the document length. The aim of the paper is to investigate the capability of two retrieval models based on the length of the query. The methodology is divided into two parts, the first shows how different term weighting schemes affects the relevance of retrieved Tweets while the second discusses how documents are annotated and split into sentences level and then used for pseudo-relevant feedback.

## B. RELATED WORK

Chun et al., [3] proposed an indexing and ranking scheme for supporting real-time search in microblogging systems that indexed only a new tweet if only it appears in the top-K results of some cached queries with high probability. Jimmy et al., [4] proposed that requesting a user include temporal information will yield more relevant result. Zhunchen et al., [5] proposed exploiting the structural information of tweets as each block encodes a variety of communicative intent and a sequence of these blocks captures changing discourse. Busch et al., [6] suggested that the requirements for a real-time system should include low-latency, high-throughput query evaluation; High ingestion rate and immediate data availability; Concurrent reads and write and dominance of the temporal signals. A commonly used query expansion technique is the Relevance Model (RM) [7]. RM is a pseudo-relevance feedback approach that makes use of retrieved documents to estimate query topic. Expansion terms that are relevant are extracted and presented based on relevance ranking and then are used in combination with the original query. RM modelled a generalized notion of relevance instead of explicitly modelling the relevance or pseudo-relevant documents as shown below.

$$P(w|Q) = \int_D P(w|D)P(D|Q)$$

$$\approx \frac{\sum_{D \in R_Q} P(w|D)P(Q|D)P(D)}{\sum_w \sum_{D \in R_Q} P(w|D)P(Q|D)P(D)} \quad \ldots(i)$$

Where $R_Q$ is the set of documents that are relevant to the query. After the model is estimated, the $k$ terms with highest likely terms from $P(.|Q)$ are extracted and interpolated with the original query expansion terms. This model forms the basis for Latent Concept Expansion (LCE) proposed Metzler et al [8], an approach that models query term dependence and achieves better retrieval effectiveness compared to the relevance model. It is based on the Markov random field model [9]. It is shown below:

$$P(E|Q) \approx \frac{\sum_{D \in R_Q} P(Q,E,D)}{\sum_E \sum_{D \in R_Q} P(Q,E,D)} \quad \ldots(ii)$$

Where $P(Q, E, D)$ is the joint probability and $R_Q$ is the set of relevant documents for the query Q. After the conditional probability has been gotten, the latent concepts E with the highest probability are picked as expansion concepts. He et al., [10] proposed an approach to detect good feedback documents thereby improving expansion effectiveness at document level. They classified all feedback documents using a variety of features such as the distribution of query terms in the feedback document, the similarity between a single feedback document and all top-ranked documents.

Onifade et al., [11] proposed a fuzzy string-matching that assist in information retrieval by guiding against the risk accruable from some class of dirty data which include strings that are misspelt, inconsistent entries, incomplete context, different ordering and ambiguous data resulting from incoherent user queries brought about by leaving the users' information need at the mercy of the system.

Hao et al., [12] proposed the model hierarchical markov random fields (HMRFs) based on LCE. Their approach modelled the various types of dependencies that exist between the original query and expansion terms. HMRF incorporated hierarchical interactions by exploiting implicit hierarchical structure within the documents.

This research work propose a robust dependency model that depicts the dependency existing amongst terms that are in a document and the effects of term weighting schemes on tweets.

## C. PROPOSED METHODOLOGY

We propose a methodology that initiates the retrieval model based on the length of query issued. The first method is applied real-time tweets gotten from Twitter, the online social networking service and the second method to RCV1, a dataset containing 800,000 reuters (document dataset). In IR, documents can be of different types and lengths, ranging from tweets, web pages, text, images, video or audio documents.

The query, *q,* submitted is first parsed into a query scanner in order to determine the retrieval model used.

Algorithm 1: *QueryScanner(q)*
*Query Length = $l_q$*
*Standard Tweet Length = $l_t$*
*If ($l_q$  $l_t$)*
    *Then TweetRetrieval(q)*
*Else*
    *SLRS(q)*

*//where SLRS is Sentence-length retrieval system*

For queries with short length of about three words or less, the Tweet Retrieval Method (I) is used as the query terms are treated like tweets while longer queries are regarded as though they are complete sentences and for this, Sentence-Length Retrieval Method (II) is used.

## I. TWEET RETRIEVAL

Bernard [13] suggested that the problem with real-time search is defining an effective ranking function. Hence, we define a ranking function that combines the term weight, timestamp and the popularity of tweet.

$$ =  +  + \quad \ldots \text{(iii)}$$

 = Ranking Function
 = Term Weight
 = Timestamp
 = Popularity of Tweet

Figure 1 shows the generic work flow for the proposed methodology. The work is divided into three parts: Query Pre-Processing, Indexing and Term Weighting.

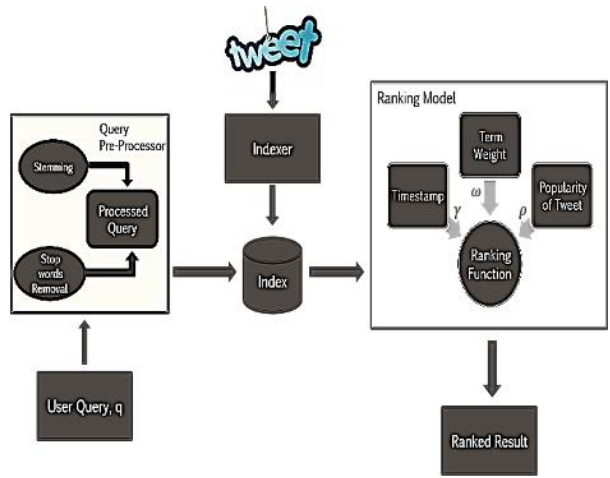

Fig 1: Proposed methodology for tweet retrieval model

## Query Pre-Processing

To have a more effective vector matching operation between query terms and documents, the query was pre-processed by performing two operations; stemming and stop words removal. Stemming brings each query term down to its root words e.g. changing the word "searching" to its root word "search" and stop words removal removes common words like prepositions (in, at, of etc.), conjunctions (and, or, with etc.) and articles (a, an, the etc.) from the search query.

Algorithm 2: *QueryPreprocessing(q)*
*Stop words, $S = \{s_i \mid i=1, 2, \ldots, n\}$*
*while(q contains $s_i$)*
   *remove $s_i$*
*Stem q to root words*
*Return preprocessed query, $q_p$*

## Indexing

When the tweets are pulled from Twitter using the Twitter streaming API, an inverted index of the tweet was created. The index built thus contained the tweet id, terms in the tweets, timestamp amongst other things.

## Term Weighting

Usually, the three main components that affect the importance of a term in a text are the term frequency (TF), Inverse Document Frequency (IDF) and document length normalization [14]. Studies have shown that good retrieval performance is closely related to the use of retrieval heuristic such as TF-IDF weighting and document length normalization, it is often necessary to modify it to achieve optimal retrieval performance [15].

A term-weighting strategy that return the most relevant tweets by employing the TF-IDF ranking is proposed. To compute the term weight, a vector matching operation that gives the similarity between the query terms and document is performed. The term weight is given as:

$$ =  \times _j \times _k \quad \ldots \text{(iv)}$$

$\alpha$ = Term Frequency
$\beta$ = Inverse Document Frequency
$\sigma$ = normalization function

*Term Frequency* shows the number of times a term *i* occurs in a document *t*. It gives preference to common words

$$= _{i,t} \qquad \text{...(v)}$$

Where $_{i,t}$ is the function depicting the frequency of term *t* in document *i*.

*Inverse Document Frequency* is used to scale down the effect of common occurring words in the collection, thereby assigning more weights to distinctly occurring words. Defined as:

$$_1 = \log\left(\frac{D}{d_i}\right) \qquad \text{...(vi)}$$

$$_2 = \log d_i \qquad \text{...(vii)}$$

Where D is the total number of documents in the collection and $d_i$ is the number of documents containing the term *i*.

*The Normalization Function* is used to make up for the effect of document length due to the fact that longer document will have more terms thereby yielding a higher TF value. We make use of cosine, pivoted cosine and pivoted unique normalization in eqns (viii) to (x) respectively. The pivoted normalization is a technique that retrieves document of all lengths with similar chances by reducing the gap between relevance and retrieval probabilities [16].

$$_1 = \frac{1}{\sqrt{\sum_{r=1}^{n}(a \times \beta_j)^2}} \qquad \text{...(viii)}$$

$$_2 = \frac{1}{(1.0 - \text{slope}) \times \text{pivot} + \text{slope} \times l} \qquad \text{...(ix)}$$

$$_3 = \frac{1}{(1.0 - \text{slope}) \times \text{pivot} + \text{slope} \times m} \qquad \text{...(x)}$$

$l$ = old normalization function
$m$ = number of unique terms

At each iteration of equation (iv), six different methods to calculate the term weight is generated.

**Popularity of Tweets**

A tweet is said to be popular based on the number of retweets, quotes and replies it has.

**Timestamp**

The weight of the timestamp is assigned based on the time the tweet is submitted, i.e., newer tweets carry more weight than older tweets. So that the results generated would be relevant and recent.

**Ranked Result**

The ranked result is presented to the user, by combining the factors discussed in eqn (iii) above in order to have tweets that are not only relevant but relevant and recent.

Algorithm 3: *TweetRetrieval(q)*
*$q_P$ = QueryPreProcessing(q)*
*Match $q_P$ with Indexed Tweets*
*Filter search using Ranking function,*
*Format Result*
*Output Result*

Higher Relevance Weight denotes Higher Rank
Where Relevance weight = | |

## II. SENTENCE-LENGTH RETRIEVAL

When a user query is long enough to be called a sentence, the IR system makes use of the sentence level expansion techniques.

The collected dataset are organized into some topic classes in a data repository after which they are each annotated and their terms grouped into sentences and extracted before a reclassified into a new repository of similar topic classes. This section emphasizes on document representation in a different format; away from the usual bag of words style. It shows term dependencies among document terms by representing this terms as sentences extracted from the pool of document dataset. The following subsection describes the different stages used to prepare and create the retrieval model. Document pre-processing, Annotation.

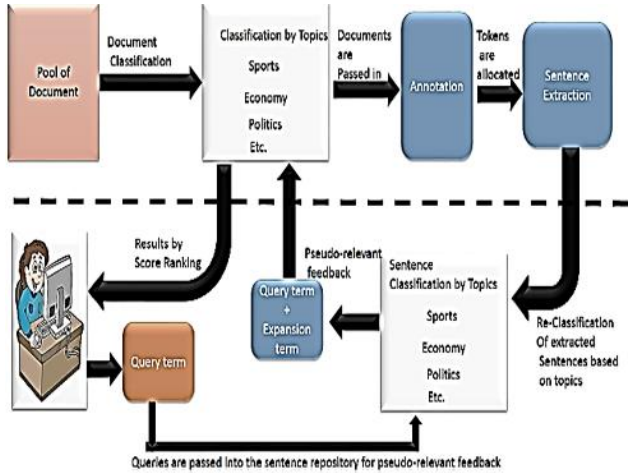

Fig 2: Proposed methodology for sentence-length retrieval model

**Document Pre-processing**

After the documents have been classified as shown in the model above. They are annotated into sentence entities. The pre-processing stage is necessary in order to make the document appear sensible. This process involved stop word removal, splitting the documents into sentences and tokenizing the sentences.

**Sentence Reclassification**

Each sentence is identified as an individual entity followed by a reclassification for the sentence repository, done by using the similar classes employed for classification of documents for the re-classification of the sentence repository. Such that

$$d_i = \sum S_1, S_2, \ldots, S_i \qquad \ldots(xi)$$

Where $S_i$ is a sentence in document $d_i$. The extracted sentences after reclassification can belong to one or more topic classes. As shown below

$$S_i \in T_1, \ldots, T_j \qquad \ldots(xii)$$

Where $T_j$ is a topic in the topic classes.

**Topic Classification**

Each topic has a method that is tagged with a set of keyword representing it rule. Each sentence is looped across this method. Every sentence whose term matches a keyword tag is added to that topic.

Algorithm 4: *topicClassification($S_i$)*
   *For each $S_i$*
      *For each $T_j$*
         *If $S_i$ contains similar terms as $T_j$*
            *Add $S_i$ to $T_j$ (include this sentence to the topic class)*

**Ranking**

The ranking was based on some features which includes term frequency model (tf), inverse document frequency (idf), and term proximity, amongst others. With these features we were able to show term dependence.

**Term Dependency**

Matching query terms with document comes with a few rules. First, terms that appear consecutively between the query and documents indicates stronger relevance. Second, a fixed proximity between the terms in both query and documents indicates strong relevance, i.e. if the proximity between two terms in a document matches the proximity between the same terms in query, then the document is relevant. The proximity and order can be achieved through the use of term dependency [17]. Using Markov random field to show dependencies for document term nodes can be overly complex because of their length but very flexible. To reduce this complexity, the document node is reduced and represented as sentence level.

**Query Expansion**

The classified sentence entities are used in the query expansion process. We make use of the Latent Concept Expansion model by Metzler et al [8] for the query expansion. The LCE model went beyond unigrams making use of entity identifiers. In LCE, the retrieval score represents the joint probability of the documents under a given query i.e. $(D|Q)$, then the document-wise multinomial distribution over a vocabulary i.e. $p(V|D)$. Combining both

$$p(V|D) = \sum_d p(V|d)p(D=d|Q) \qquad \ldots(xiii)$$

**Features**

Features are the backbone to information retrieval. Term frequency (tf), term proximity and inverse document frequency (idf) are the most used features. The major differences in most information retrieval models are how these features are combined.

Algorithm 5: *SLRS(q)*
$S_e$ = *QueryExpansion(q)*
//where $S_e$ represents expanded query
//Document retrieval
For each $S_e$
    add *termProximityFeature($S_e$)*
    add *tf.idfFeature($S_e$)*
    *QueryPreprocessing($S_e$)*
If $S_e$ is an element of $d_i$ then
    Return $d_i$ by ranking score;

Algorithm 6: *SentenceExtrtaction($d_i$)*
*Annotate($d_i$)*
    *ExtractSentences($S_1, S_2 ... S_i$) from $d_i$*
For each $S_i$
    *topicClassification($S_i$)*

Algorithm 7: *QueryExpansion(q)*
    add *termProximityFeature(q)*
    add *tf.idfFeature(q)*
    *QueryPreprocessing(q)*
    if terms in q are a subset of $S_i$ then
        return $S_i$ by ranking score
        $q_i + S_i = S_e$ (expanded query)

### D. EXPERIMENT AND CONCLUSION

With the above algorithms and mathematical models, we conclude that the proposed retrieval system would produce a more effective and efficient retrieval for tweet and sentence length queries. We hope that in the future of this work, we would implement the framework in order to have a more conclusive result.

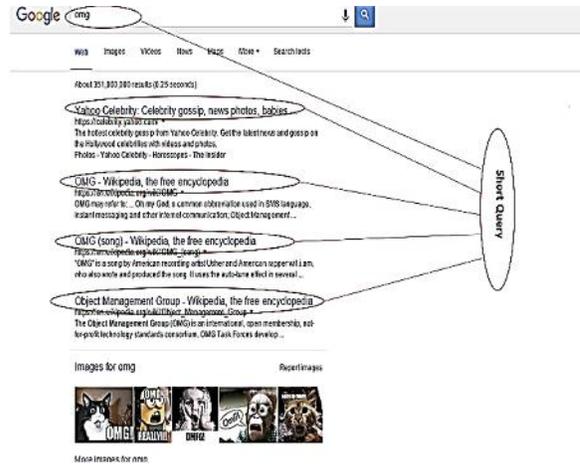

Fig 3: Short Query Evaulation on Google

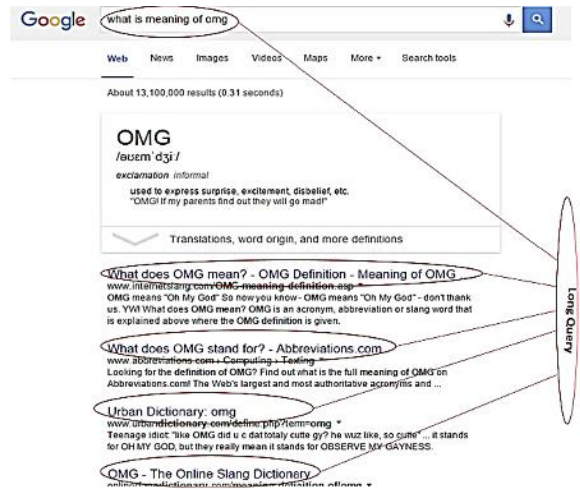

Fig 4: Long Query Evaulation on Google

From the above query search on "omg", we see that longer queries retrieves more specific and precise documents and very short queries retrieve results that are diverse and less specific. Tweets are very shorts and may contain sentiments attached by the users. This shows the importance of query length in a retrieval system as observed in Algorithm 1. Google search engine represents all queries as bag of words and as such, very short queries results may not contain a user's document choice. As observed in fig 4, the user query was more detailed which led to a more detailed result.